\documentclass[prd,reprint,nofootinbib,superscriptaddress,preprintnumbers,showpacs]{revtex4-1}
\usepackage{amsfonts}
\usepackage{amssymb}
\usepackage{amsmath}
\usepackage{graphicx,color}
\usepackage{dcolumn}
\usepackage{epsfig}
\usepackage{bm}
\usepackage{bbm}
\usepackage{ulem}
\usepackage{hyperref}
\usepackage{lineno}

\usepackage{color}

\def\gtap{\ \raise.3ex\hbox{$>$\kern-.75em\lower1ex\hbox{$\sim$}}\ }
\def\ltap{\ \raise.3ex\hbox{$<$\kern-.75em\lower1ex\hbox{$\sim$}}\ }
\begin{document}

\title{
$P_c(4312)^+$ and $P_c(4337)^+$ as interfering $\Sigma_c\bar{D}$ and $\Lambda_c\bar{D}^{*}$
threshold cusps 
}
\author{S.X. Nakamura}
\email{satoshi@ustc.edu.cn}
\affiliation{
University of Science and Technology of China, Hefei 230026, 
People's Republic of China
}
\affiliation{
State Key Laboratory of Particle Detection and Electronics (IHEP-USTC), Hefei 230036, People's Republic of China}
\author{A. Hosaka}
\affiliation{
Research Center for Nuclear Physics, Osaka University, 
Ibaraki 567-0047, Japan
}
\affiliation{
Advanced Science Research Center, Japan Atomic Energy Agency (JAEA), Tokai 319-1195, Japan 
}
\author{Y. Yamaguchi}
\affiliation{
Advanced Science Research Center, Japan Atomic Energy Agency (JAEA), Tokai 319-1195, Japan 
}

\begin{abstract}
The recent LHCb data on 
$B_s^0\to J/\psi p\bar{p}$
revealed a new pentaquark-like 
$P_c(4337)^+$ structure,
while finding no evidence for 
$P_c(4312)^+$ discovered earlier in
$\Lambda_b^0\to J/\psi p K^-$.
Though puzzling, the data 
actually offer an important hint to understand the nature of the
pentaquark candidates. 
We develop a model to analyze the $B_s^0\to J/\psi p\bar{p}$ data.
We find that 
a $\Sigma_c\bar{D}$ one-loop mechanism causes 
a threshold cusp that fits well the $P_c(4337)^+$ peak.
Also, the  $\Sigma_c\bar{D}$ and $\Lambda_c\bar{D}^{*}$ 
threshold cusps interfere with each other to reproduce 
an oscillating behavior in 
the proton helicity angle distribution.
These results combined with our earlier analysis on 
$\Lambda_b^0\to J/\psi p K^-$
 indicate that
$P_c(4312)^+$ and
$P_c(4337)^+$
are created by different interference patterns between 
the $\Sigma_c\bar{D}$ 
and $\Lambda_c\bar{D}^{*}$ 
(anomalous) threshold cusps.
The proposed scenario consistently explains why
the $P_c(4312)^+$ and $P_c(4337)^+$ peaks appear 
in $\Lambda_b^0\to J/\psi p K^-$ and $B_s^0\to J/\psi p\bar{p}$,
respectively, but not vice versa or both. 
\end{abstract}

\maketitle

\section{introduction}

Since the discovery of $X(3872)$~\cite{belle_x3872_jpsi-rho},
we have witnessed
many experimental observations of 
exotic hadron candidates
that are not categorized in the conventional constituent quark
structures such as $q\bar{q}$ and $qqq$;
see Refs.~\cite{review_chen,review_hosaka,review_lebed,review_esposito,review_ali,review_guo,review_olsen,review_Brambilla}
for reviews.
Yet, we have not reached a satisfactory
understanding of their nature.  
Many theoretical attempts have been made to interpret the candidates
as hadronic molecules~\cite{pc_beihang,pc_valencia,pc_beihang2,pc_hebei,pc_nanjin,pc_itp,pc_chen,pc_lanzhou,pc_wang,pc_gutsche,pc_peking,pc_nanjing2,pc_lin,pc_burns,pc_xu,pc_yamaguchi,pc_sakai,pc_voloshin,pc_wu,pc_jrzhang,pc_hxu,pc_du,pc_du2,pc_xiao,pc_ling,pc_xling,pc_mwli,Azizi:2020ogm,Yamaguchi:2017zmn,Yamaguchi:2016ote}, 
compact multiquarks~\cite{pc_ali,pc_pimikov,pc_zgwang,pc_rzhu,pc_xzweng,pc_bari,pc_stancu,pc_ydong,Santopinto:2016pkp},
or hadrocharmonia~\cite{Ferretti1,Ferretti2,pc_hadrochamonium}.
However, difficulties of these interpretations lie in: 
(i) We poorly understand hadron interactions and thus 
formation mechanisms of the exotic hadrons;
(ii) Although resonances
are expected to appear in different processes, this is often not the
case and their appearances seem to be highly process-dependent.

Kinematical effects, (anomalous) threshold cusps in particular, are
an interesting alternative mechanism that can create resonancelike
structures~\cite{ts_review}.
They are rather free from the above item (i)
since their singular behaviors occur when certain kinematical conditions
are satisfied, irrespective of details of dynamics. 
The kinematical conditions are often satisfied in one process but not in
others and, thus, 
the item (ii) is naturally understood.
In this Letter, 
taking advantage of these characteristics of the kinematical effects, we propose a scenario
on
how one can consistently explain
pentaquark candidates $P_c$'s 
observed in
the recently measured 
$\Lambda_b^0$ and $B_s^0$ decay processes.

The pentaquark candidates were first discovered
in $\Lambda_b^0\to J/\psi pK^-$~\footnote{
We follow the hadron naming scheme of~\cite{pdg}. 
For simplicity, however, 
$\Sigma_c(2455)^{+(++)}$,
$\Sigma_c(2520)^{+(++)}$, and $J/\psi$ are often denoted 
by $\Sigma_c$, $\Sigma^*_c$,
 and $\psi$, respectively.
We also collectively denote
$\Lambda_c$ and $\Sigma_c$ by $Y_c$.
Charge indices are often suppressed.
}
by the LHCb Collaboration~\cite{lhcb_pc_old}.
The recent update with $\sim 10$ times larger statistics found 
in the $J/\psi p$ invariant mass ($M_{J/\psi p}$) distribution
three clear peaks that were assigned to pentaquarks called
$P_c(4312)^+$, $P_c(4440)^+$, and $P_c(4457)^+$~\cite{lhcb_pc}.
To establish $P_c^+$'s as hadronic states, 
their appearances in different processes 
have been studied~\cite{photo_qwang,photo_kubarovsky,photo_Karliner,photo_hiller,photo_xywang,photo_wu,photo_cao}.
However, the current experimental situation is not very supportive in
this regard,
though higher statistics data in the future might change the situation.
For instance, 
a $P_c$ search in 
the GlueX $J/\psi$ photoproduction cross section measurement
ended up with a null result~\cite{gluex}.
Moreover, in $\Lambda_b^0\to J/\psi p \pi^-$ data from the LHCb,
an enhanced number of events is seen in
the $M_{J/\psi p}$ bin of $P_c(4440)^+$
while this is not the case for 
the other $P_c^+$'s~\cite{Pc_lhcb2}.
Very recent LHCb data on $B_s^0\to J/\psi p\bar{p}$
has made the situation even more complex~\cite{lhcb_pc4337}.
Their four-dimensional amplitude analysis found evidence for a new $P_c(4337)^+$ signal
with a $3.1-3.7~\sigma$ significance, while no evidence was found for
$P_c(4312)^+$.~\footnote{
We do not discuss $P_c(4440)^+$ and $P_c(4457)^+$ hereafter since
they are located (virtually) outside of the
phase-space of $B_s^0\to J/\psi p\bar{p}$.}
Recently, one of the present authors 
showed that the 
$P_c(4312)^+$ peak in $\Lambda_b^0\to J/\psi pK^-$
can be formed by an interference between 
kinematical effects, including a novel double triangle singularity (DTS), and
a smooth amplitude~\cite{DTS,DTS-pos}.
Within this picture, 
we can explain 
the absence of 
$P_c(4312)^+$ in the other processes since
the DTS does not exist and the interference pattern is different.
A similar idea can also be applied 
to the $P_c(4337)^+$ signal in $B_s^0\to J/\psi p\bar{p}$.

In this work, 
we develop a model for
$B_s^0\to J/\psi p\bar{p}$ that consists of
one-loop [Fig.~\ref{fig:diag}(a)]
and direct decay [Fig.~\ref{fig:diag}(b)] mechanisms.
The one-loop amplitudes cause threshold cusps in the 
$M_{J/\psi p}$ distribution.
We demonstrate that 
the $\Sigma_c\bar{D}$ threshold cusp develops a 
structure that fits well the $P_c(4337)^+$ peak.
Furthermore, 
the $\Sigma_c\bar{D}$ and $\Lambda_c\bar{D}^*$ one-loop amplitudes, 
which are expected to have comparable magnitudes, interfere with each
other to create
an oscillating behavior in the proton helicity angle distribution in a
remarkable agreement with the data.
With these results, we provide a reasonable explanation 
why 
the resonancelike
 $P_c(4312)^+$ and $P_c(4337)^+$ peaks 
selectively
appear
in $\Lambda_b^0\to J/\psi pK^-$ 
and $B_s^0\to J/\psi p \bar{p}$,
respectively.

\section{model}

\begin{figure}[t]
\begin{center}
\includegraphics[width=.48\textwidth]{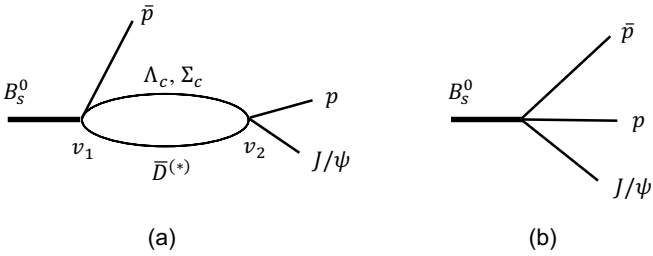}
\end{center}
 \caption{
$B_s^0\to J/\psi p \bar{p}$ mechanisms considered in this work:
(a) one-loop;
(b) direct decay.
Diagrams with charge-conjugated 
intermediate and final states
are also considered.
 }
\label{fig:diag}
\end{figure}

\begin{figure}[b]
\begin{center}
\includegraphics[width=.48\textwidth]{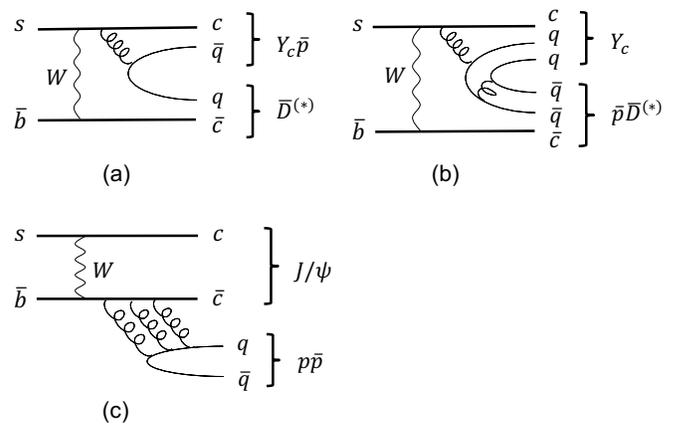}
\end{center}
 \caption{
Possible quark diagrams for (a,b) $B_s^0\to Y_c\bar{p}\bar{D}^{(*)}$
and (c) $B_s^0\to J/\psi p\bar{p}$; $q=u$ or $d$.
 }
\label{fig:quark}
\end{figure}

The one-loop mechanisms [Fig.~\ref{fig:diag}(a)] are initiated by
$B^0_s\to Y_c\bar{D}^{(*)}(J^P)\bar{p}$ with
$Y_c\bar{D}^{(*)}(J^P)=\Lambda_c\bar{D}^0(1/2^-)$,
$\Lambda_c\bar{D}^{*0}(1/2^-)$, and
$\Sigma_c\bar{D}(1/2^-)$;
$J^P$ stands for the spin ($J$) and parity ($P$) of the 
$Y_c\bar{D}^{(*)}$ pair.
These weak decays would be mainly caused by 
quark-level mechanisms shown in 
Figs.~\ref{fig:quark}(a) and \ref{fig:quark}(b).
The $Y_c\bar{p}$ in Fig.~\ref{fig:quark}(a)
could be generated from a decay chain such as 
$B_s^0\to D^{**}\bar{D}^{(*)}$, 
$D^{**}\to Y_c\bar{p}$
with $D^{**}$ being an excited charmed meson.
In this case, a $D^{**}Y_c\bar{D}^{(*)}$ triangle diagram 
would contribute to $B_s^0\to J/\psi p \bar{p}$.
However, such triangle diagrams may be simulated 
by Fig.~\ref{fig:diag}(a) because,
with experimentally confirmed charmed mesons,
they do not cause triangle singularities.
Similarly,
the diagram of Fig.~\ref{fig:quark}(b) might induce 
a $\bar{Y}_c^{**}Y_c\bar{D}^{(*)}$ 
triangle diagram
that may be simulated
by Fig.~\ref{fig:diag}(a).

Another triangle diagram can be drawn 
from Fig.~\ref{fig:diag}(a)
by replacing the contact interaction $v_2$ with 
a $D^{(*)}$-exchange mechanism.
This triangle diagram is similar to 
initial single pion emission (ISPE) mechanisms
that have been used to describe 
the $Z_b(10610)$ and $Z_b(10650)$ structures~\cite{ISPE_zb}
and the $Z_c(3900)$ structure~\cite{ISPE_zc}.
Since the exchanged $D^{(*)}$-meson is highly off-shell, this ISPE-like
mechanism should be well approximated and included in the one-loop mechanism of 
Fig.~\ref{fig:diag}(a).
The contact interaction $v_2$ would also include other possible
short-range mechanisms such as a quark-exchange.

Regarding the direct decay [Fig.~\ref{fig:diag}(b)],
we expect and assume 
it to be also initiated by
the quark diagrams of Figs.~\ref{fig:quark}(a) and \ref{fig:quark}(b).
Thus, the direct decay simulates 
one-loop mechanisms including heavier $Y_c\bar{D}^{(*)}(J^P)$ such as 
$\Sigma_c\bar{D}^*(1/2^-)$.
Meanwhile,
one may expect another quark diagram of 
Fig.~\ref{fig:quark}(c) as a main driver for
$B^0_s\to J/\psi p\bar{p}$~\cite{lhcb_2019}.
However, 
the $p\bar{p}$ pair created in this way must go through a strong final
state interaction, causing a significant $p\bar{p}$ threshold enhancement~\cite{ppbar-model,ppbar-model2}.
The absence of such enhancement in the LHCb data~\cite{lhcb_pc4337}
would indicate that 
this quark diagram
plays a minor role.

Now we present formulas for the one-loop amplitudes.
Masses and widths of the particles that contribute to the amplitudes are 
taken from Ref.~\cite{pdg}.
We denote the energy, width, momentum, and polarization vector of a particle $x$ by
$E_x$, $\Gamma_x$, $\bm{p}_x$, and $\bm{\epsilon}_x$, respectively.
The initial 
$B^0_s\to Y_c\bar{D}^{(*)}(J^P)\bar{p}$ vertices with
$Y_c\bar{D}^{(*)}(J^P)=\Lambda_c\bar{D}^0(1/2^-)$,
$\Lambda_c\bar{D}^{*0}(1/2^-)$, and
$\Sigma_c\bar{D}(1/2^-)$
are given by
\begin{eqnarray}
\label{eq:dt11}
v_{1a}&=& c^{1/2^-}_{\Lambda_c \bar{D}\bar{p},B_s}\,
 F_{\Lambda_c \bar{D} \bar{p},B_s}^{00}\ , \\
v_{1b}&=& c^{1/2^-}_{\Lambda_c \bar{D}^*\bar{p},B_s}\,
\bm{\sigma}\cdot\bm{\epsilon}_{\bar{D}^*}
 F_{\Lambda_c \bar{D}^* \bar{p},B_s}^{00}\ , \\
v_{1c}&=& c^{1/2^-}_{\Sigma_c \bar{D}\bar{p},B_s}\,
\bigg(1 t_{\Sigma_c} {1\over 2} t_{\bar{D}}\bigg| {1\over 2}{1\over 2}\bigg)
 F_{\Sigma_c \bar{D} \bar{p},B_s}^{00}\ , 
\label{eq:dt12}
\end{eqnarray}
respectively,
with complex  coupling constants 
$c^{J^P}_{Y_c\bar{D}^{(*)}\bar{p},B_s}$.
The parenthesis is an isospin Clebsch-Gordan coefficient with 
$t_x$ being the isospin $z$-component of a particle $x$.
We have used dipole form factors $F_{ijk,l}^{LL'}$ defined by
\begin{eqnarray}
\label{eq:ff1}
 F_{ijk,l}^{LL'} =
 {1\over \sqrt{E_i E_j E_k E_l}}
\left(\frac{\Lambda^2}{\Lambda^2+q_{ij}^2}\right)^{\!\!2+{L\over 2}}\!\!\!
\left(\frac{\Lambda^{2}}{\Lambda^{2}+\tilde{p}_k^2}\right)^{\!\!2+{L'\over 2}}\!\!\!\!\!\!,
\end{eqnarray}
where $q_{ij}$ ($\tilde{p}_{k}$) is the momentum of $i$ ($k$) in the
$ij$ (total) center-of-mass frame.
Unless otherwise stated,
we use a common cutoff value $\Lambda=1$~GeV 
in Eqs.~(\ref{eq:ff1}) and (\ref{eq:ff2})
 for all the interaction vertices.
The subsequent $s$-wave interactions for 
$\Lambda_c^{+}\bar{D}^{0}(1/2^-)$,
$\Lambda_c^{+}\bar{D}^{*0}(1/2^-)$,
$\Sigma_c\bar{D}(1/2^-)\to J/\psi p$
are given by separable interactions
 with coupling constants
$c^{J^P}_{\psi p, Y_c\bar{D}^{(*)}}$ as
\begin{eqnarray}
v_{2a}&=& c^{1/2^-}_{\psi p, \Lambda_c\bar{D}}
\bm{\sigma}\cdot \bm{\epsilon}_\psi \,
 f_{\psi p}^{0}
 f_{\Lambda_c\bar{D}}^{0} \ , \\
v_{2b}&=& c^{1/2^-}_{\psi p, \Lambda_c\bar{D}^*}
\bm{\sigma}\cdot \bm{\epsilon}_\psi \,
\bm{\sigma}\cdot\bm{\epsilon}_{\bar{D}^*}
 f_{\psi p}^{0}
 f_{\Lambda_c\bar{D}^*}^{0} \ , \\
v_{2c}&=& c^{1/2^-}_{\psi p, \Sigma_c\bar{D}}
\bigg(1 t_{\Sigma_c} {1\over 2} t_{\bar{D}}\bigg| {1\over 2}{1\over 2}\bigg)
\bm{\sigma}\cdot \bm{\epsilon}_\psi \,
 f_{\psi p}^{0}\,
 f_{\Sigma_c\bar{D}}^{0}  , 
\end{eqnarray}
with form factors
\begin{eqnarray}
 f_{ij}^{L} =
 {1\over \sqrt{E_i E_j}}
\left(\frac{\Lambda^2}{\Lambda^2+q_{ij}^2}\right)^{2+(L/2)}\ .
\label{eq:ff2}
\end{eqnarray}
The one-loop amplitudes including $Y_c\bar{D}^{(*)}(J^P)$,
denoted by $A^{\rm 1L}_{Y_c\bar{D}^{(*)}(J^P)}$,
are given with the above ingredients as
\begin{widetext}
\begin{eqnarray}
\label{eq:1L2}
A^{\rm 1L}_{\Lambda_c\bar{D}(1/2^-)} &=&
 c^{1/2^-}_{\psi p, \Lambda_c\bar{D}}\,
 c^{1/2^-}_{\Lambda_c \bar{D}\bar{p},B_s}\,
\bm{\sigma}\cdot \bm{\epsilon}_\psi \,
\int d^3p_{\bar{D}}
{ 
 f_{\psi p}^{0}
 f_{\Lambda_c\bar{D}}^{0} 
 F_{\Lambda_c \bar{D} \bar{p},B_s}^{00}
\over W-E_{\Lambda_c}-E_{\bar{D}}
+i\epsilon
}\ ,
\\
\label{eq:1L4}
A^{\rm 1L}_{\Lambda_c\bar{D}^*(1/2^-)} &=&
3\,c^{1/2^-}_{\psi p, \Lambda_c\bar{D}^*}\,
c^{1/2^-}_{\Lambda_c \bar{D}^*\bar{p},B_s}\,
\bm{\sigma}\cdot \bm{\epsilon}_\psi \,
\int d^3p_{\bar{D}^*}
{ 
 f_{\psi p}^{0}
 f_{\Lambda_c\bar{D}^*}^{0}
 F_{\Lambda_c \bar{D}^* \bar{p},B_s}^{00}
\over W-E_{\Lambda_c}-E_{\bar{D}^*}
+i\epsilon
}\ ,
\\
A^{\rm 1L}_{\Sigma_c\bar{D}(1/2^-)} &=&
 c^{1/2^-}_{\psi p, \Sigma_c\bar{D}}\,
 c^{1/2^-}_{\Sigma_c \bar{D}\bar{p},B_s}\,
\bm{\sigma}\cdot \bm{\epsilon}_\psi \,
\sum_{\rm charge}
\bigg(1 t_{\Sigma_c} {1\over 2} t_{\bar{D}}\bigg| {1\over 2} {1\over 2}\bigg)^2
\int d^3p_{\bar{D}}
{ 
 f_{\psi p}^{0}\,
 f_{\Sigma_c\bar{D}}^{0}
 F_{\Sigma_c \bar{D} \bar{p},B_s}^{00}
\over W-E_{\Sigma_c}-E_{\bar{D}}
+{i\over 2}\Gamma_{\Sigma_c}
}\ ,
\label{eq:1L3}
\end{eqnarray}
\end{widetext}
where $\sum_{\rm charge}$ indicates 
the summation over 
$\Sigma_c^{+} \bar{D}^0$ and
$\Sigma_c^{++} D^-$
intermediate states
with the charge dependent particle masses;
$W\equiv E-E_{\bar{p}}$;
the tiny $\Gamma_{D^*}$ has been neglected.
The above expressions are implicitly sandwiched by the spinors of the
final $p$ and $\bar{p}$.

The one-loop amplitudes in Eqs.~(\ref{eq:1L2})-(\ref{eq:1L3}) are
further supplemented by 
$Y_c\bar D^{(*)}$ scattering amplitudes.
Here, we use 
a $Y_c\bar D^{(*)}$ single channel scattering model used in Ref.~\cite{DTS};
see Sec.~2 of the Supplemental Material in
Ref.~\cite{DTS} for details. 
We assume
possible coupled-channel and/or multi-loop effects to be absorbed by the 
coupling constants that are  fitted to the data. 
The $Y_c\bar D^{(*)}$ interaction strengths are fixed such that
the scattering lengths~\footnote{
The scattering length $(a)$ is related to the phase shift $(\delta)$ by 
$p\cot\delta=1/a + {\cal O}(p^2)$.} 
are $a \sim 0.5$~fm (attractive) for $\Sigma_c\bar D$
while $a \sim -0.2$~fm (repulsive) for $\Lambda_c\bar D^*$
as in Ref.~\cite{DTS}.
Regarding $\Lambda_c\bar D$ not considered in Ref.~\cite{DTS}, 
we use a scattering amplitude of $a \sim -0.2$~fm
as the fit prefers a repulsion.
The one-loop amplitudes create threshold cusps in the $M_{J/\psi p}$ spectrum 
that become significantly sharper [less sharp]
as the $Y_c\bar{D}^{(*)}$ interactions are more attractive [repulsive]~\cite{xkdong}.
However, 
the positions of the cusps are 
not very sensitive to the $a$ values
since they are determined by the kinematical effects.

For the direct decay amplitude, we employ 
the following one for $p J/\psi(1/2^-)$ partial wave as
\begin{eqnarray}
\label{eq:dir_s}
 A_{\rm dir} &=&
  c_{\rm dir}^{1/2^-}
\bm{\sigma}\cdot \bm{\epsilon}_\psi  \,
 F_{p\psi\bar{p},B_s}^{00} \ ,
\end{eqnarray}
where $c_{\rm dir}^{1/2^-}$ is a coupling constant.
This amplitude creates $J/\psi p\bar{p}$ 
such that all two-particle pairs are in $s$-wave
and, therefore, 
it has a similarity to 
the non-resonant
amplitude used in the LHCb analysis~\cite{lhcb_pc4337}.

Our full $B_s^0\to J/\psi p\bar{p}$ decay amplitude also includes the 
diagrams obtained from Fig.~\ref{fig:diag} by replacing 
the intermediate and final particles with their charge-conjugates. 
The amplitudes for such charge analogous mechanisms
are obtained from Eqs.~(\ref{eq:1L2})-(\ref{eq:dir_s}) by 
replacements $Y_c\to \bar{Y}_c$ and $\bar{D}^{(*)}\to D^{(*)}$,
and by interchanges $p\leftrightarrow \bar{p}$,
assuming that the coupling constants do not change.
Thus, the Dalitz plot distribution from this model is
symmetric with respect to interchanging $p$ and $\bar{p}$.
See Appendix~B of Ref.~\cite{3pi} for details on how to calculate 
the Dalitz plot distribution using the presented amplitudes.

\section{results}

\begin{table}[b]
\renewcommand{\arraystretch}{1.6}
\tabcolsep=4.mm
\caption{\label{tab:para3}
Parameter values for our $B_s^0\to J/\psi p\bar{p}$ models.
The second (third) column is for 
the model that includes (does not include) the single channel $Y_c\bar{D}^{(*)}$ scattering amplitudes.
The parameters above (below) the horizontal line are 
obtained from (not) fitting the LHCb data~\cite{lhcb_pc4337}.
The parameters above the horizontal line can be arbitrarily scaled by 
a common complex overall factor. 
The last three parameters are $Y_c\bar{D}^{(*)}\to Y_c\bar{D}^{(*)}$
 interaction strengths defined by
Eq.~(43) in the Supplemental Material of Ref.~\cite{DTS}.
}
\begin{tabular}{lrr}
$ c^{1/2^-}_{\psi p, \Lambda_c\bar{D}}\, c^{1/2^-}_{\Lambda_c \bar{D}\bar{p},B_s}$   &$  0.25 -1.48 \,i$&$ -0.09 -0.71 \,i$   \\
$c^{1/2^-}_{\psi p, \Lambda_c\bar{D}^*}\,c^{1/2^-}_{\Lambda_c \bar{D}^*\bar{p},B_s}$ &$  1.29 -0.02 \,i$&$  0.49 +0.40 \,i$   \\
$ c^{1/2^-}_{\psi p, \Sigma_c\bar{D}}\, c^{1/2^-}_{\Sigma_c \bar{D}\bar{p},B_s}$     &$ -0.99 -1.99 \,i$&$  1.33 -4.54 \,i$  \\
$c_{\rm dir}^{1/2^-}$              &$  2.16 +0.00 \,i$&$  1.52 +0.00 \,i$ \\\hline
$\Lambda$ (MeV) &1000 &1000  \\
$h_{\Lambda_c\bar{D}(1/2^-)}$ & 2.5 & 0   \\
$h_{\Lambda_c\bar{D}^*(1/2^-)}$ & 2.5& 0  \\
$h_{\Sigma_c\bar{D}(1/2^-)}$ & $-2$ & $0$ 
\end{tabular}
\end{table}

\begin{figure*}[t]
\begin{center}
\includegraphics[width=1\textwidth]{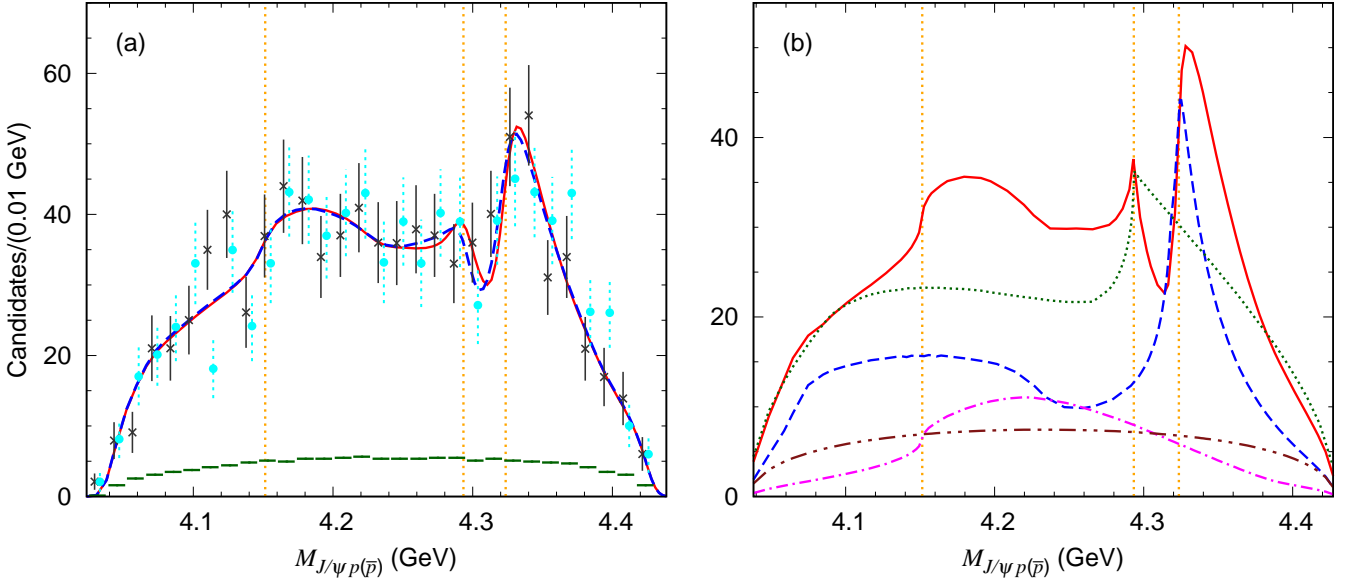}
\end{center}
 \caption{
$J/\psi p$ and $J/\psi \bar{p}$ invariant mass
 distributions for $B_s^0\to J/\psi p \bar{p}$.
(a) Comparison with the LHCb data~\cite{lhcb_pc4337}.
The red solid [blue dashed] curve is from the full model
with [without] the $Y_c\bar{D}^{(*)}$ single-channel scattering amplitudes,
augmented by experimental background (green histogram)~\cite{lhcb_pc4337}
and smeared with the bin width.
The dotted vertical lines indicate thresholds for,
 from left to right, 
$\Lambda_c^+\bar{D}^{0}$, 
$\Lambda_c^+\bar{D}^{*0}$, 
and 
$\Sigma_c(2455)^{++}D^-$,
respectively.
The black crosses (cyan circles) are
the $J/\psi p$ ($J/\psi \bar{p}$) data 
and are shifted by $-2$~MeV (+2~MeV) 
along the $M_{J/\psi p(\bar{p})}$ axis 
for better visibility.
(b) Contribution from each mechanism.
The blue dashed, green dotted, and magenta dash-dotted
curves are from the
$\Sigma_c\bar{D}$, $\Lambda_c\bar{D}^*$, and $\Lambda_c\bar{D}$ one-loop
 mechanisms, respectively;
the corresponding charge-conjugate
$\bar{Y}_c D^{(*)}$ one-loop is also included in
 each. 
The brown dash-two-dotted curve is from the direct decay mechanism.
These contributions are coherently summed to give 
the red solid curve. The curves are not smeared.
 }
\label{fig:comp-data}
\end{figure*}

Our $B_s^0\to J/\psi p \bar{p}$ model is 
fitted to the LHCb data for the
$M_{J/\psi p}$ and $M_{J/\psi \bar{p}}$ distributions 
and also for the proton helicity angle ($\theta_p$) distribution;
$\cos\theta_p\equiv -\hat{p}_p\cdot\hat{p}_{\psi}$
with $\hat{p}_x\equiv \bm{p}_x/|\bm{p}_x|$
in the $p\bar{p}$-at-rest frame.
As the LHCb data is the sum of the $B_s^0$ and $\bar{B}_s^0$ decays,
the data is symmetric with respect to interchanging $p$ and $\bar{p}$
under the $CP$ conservation. 
Regarding the fitting parameters, 
each of the one-loop amplitudes [Eqs.~(\ref{eq:1L2})-(\ref{eq:1L3})]
has  an independent complex parameter 
$c^{1/2^-}_{\psi p, Y_c\bar{D}^{(*)}}c^{1/2^-}_{Y_c\bar{D}^{(*)}\bar{p},B_s}$. 
The direct decay amplitude [Eq.~(\ref{eq:dir_s})] also has a coupling
constant $c_{\rm dir}^{1/2^-}$.
Since the overall normalization and phase of the full amplitude are arbitrary, we totally have six
fitting parameters.

A reasonable fit is obtained for the $M_{J/\psi p}$ and $M_{J/\psi\bar{p}}$
distributions as shown in Fig.~\ref{fig:comp-data}(a);
the theoretical curves have been smeared with the bin width.
By working with the six fitting parameters,
we do not find another solution of a comparable quality. 
Parameter values are given in Table~\ref{tab:para3}.
We have plotted the results for two models, one with
the $Y_c\bar{D}^{(*)}$ single channel scattering amplitude
included (red solid curve), and the other without
(blue dashed curve). 
As we observe in the figure, the fit quality is not significantly
different between the two models.
The $\Lambda_c^+\bar{D}^{0}$, 
$\Lambda_c^+\bar{D}^{*0}$, 
and 
$\Sigma_c \bar{D}$ one-loop amplitudes create 
cusp structures near their thresholds. 
In particular, 
the $\Sigma_c \bar{D}$ threshold cusp 
appearing slightly above the $\Sigma_c \bar{D}$ threshold
well describes the $P_c(4337)^+$ peak structure.
The cusp can mimic a $1/2^-$ resonance contribution, which is consistent with 
one of possible $P_c(4337)^+$ 
spin-parity assignments found in 
Ref.~\cite{lhcb_pc4337}.
Thus, within our model, 
$P_c(4337)^+$ does not exist as an exotic hadron.

Contributions from various mechanisms are plotted in 
Fig.~\ref{fig:comp-data}(b);
the $Y_c\bar{D}^{(*)}$ single channel scattering amplitudes
are implemented.
The spectra are not smeared with the bin width.
The blue dashed curve is from the coherently summed
$\Sigma_c \bar{D}$ and $\bar{\Sigma}_c D$ 
one-loop amplitudes.
The $\Sigma_c \bar{D}$ and $\bar{\Sigma}_c D$ one-loop amplitudes 
create a threshold cusp 
and a reflection (broad bump in $M_{J/\psi p}= 4.1-4.2$~GeV),
respectively, in the fairly separated $M_{J/\psi p}$
regions.
Due to the attractive $\Sigma_c \bar{D}$ interaction, the threshold cusp
is rather sharp.
Similarly, 
the green dashed curve is from 
the $\Lambda_c^+\bar{D}^{*0}$ 
and $\bar{\Lambda}_c D^{*0}$ 
one-loop amplitudes.
The repulsive $\Lambda_c^+\bar{D}^{*0}$ interaction 
creates a less sharp threshold cusp.
Meanwhile, 
the $\Lambda_c^+\bar{D}^{0}$ 
and $\bar{\Lambda}_c D^{0}$ 
one-loop amplitudes significantly 
interfere with each other, and
the $\Lambda_c^+\bar{D}^{0}$ threshold cusp does not appear as a peak
structure (magenta dash-dotted curve).

\begin{figure}[t]
\begin{center}
\includegraphics[width=.5\textwidth]{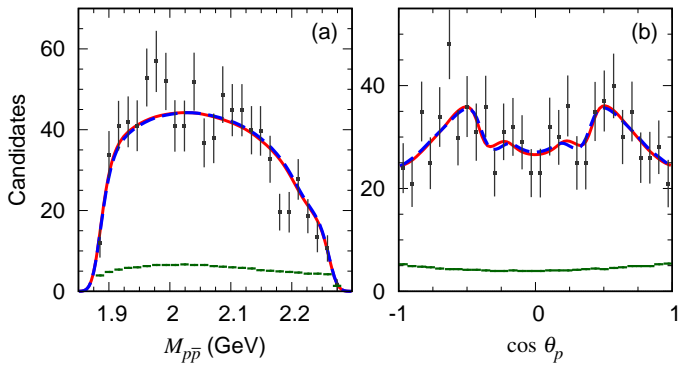}
\end{center}
 \caption{
(a) $p\bar{p}$ invariant mass
and (b) proton helicity angle
 distributions for
$B_s^0\to J/\psi p \bar{p}$.
The other features are the same as Fig.~\ref{fig:comp-data}.
 }
\label{fig:comp-data-pp}
\end{figure}

The calculation is also compared with the 
$M_{p\bar{p}}$ distribution data in 
Fig.~\ref{fig:comp-data-pp}(a).
Although this observable is not included in the fitting procedure,
 the agreement is reasonable. 

Finally, a comparison is made for the proton helicity angle distribution data
in Fig.~\ref{fig:comp-data-pp}(b).
Remarkably, our model reproduces
an oscillating behavior rather well.
The $\Sigma_c \bar{D}$ and $\bar{\Sigma}_c {D}$ threshold cusps create
the larger peaks at $\cos\theta_p\sim \pm 0.5$, and
an interference with  
the $\Lambda_c^+\bar{D}^{*0}$ and $\bar{\Lambda}_c D^{*0}$ threshold cusps
causes the smaller peaks at $\cos\theta_p\sim \pm 0.2$.
On the other hand, 
in the LHCb analysis~\cite{lhcb_pc4337}, 
while the $P_c(4337)^+$ and $\bar{P}_c(4337)$ resonant mechanisms
well reproduce
the peaks at $\cos\theta_p\sim \pm 0.5$,
their model's flat distribution in $|\cos\theta_p| \ltap 0.3$
is qualitatively different from the
seemingly oscillating data.
Yet,
the LHCb's flat distribution is still
consistent with the data within
large errors.
It is important to confirm 
the oscillating behavior 
with higher statistics data to discriminate different models.

We varied the common cutoff in the form factors over $\Lambda=0.8-2$~GeV, and refitted the
parameters. 
The fit quality does not significantly change. 
This stability is expected since the singular behaviors of the threshold
cusps do not sensitively depend on the cutoff.

Can we observe effects of 
the $Y_c\bar{D}^{(*)}$ coupled-channel dynamics in the LHCb data ?
The red solid and blue
dashed curves in Figs.~\ref{fig:comp-data}(a) and 
\ref{fig:comp-data-pp} are rather similar.
The fit quality does not significantly change as long as
the $Y_c\bar{D}^{(*)}$ scattering length is varied 
within a range 
where the sign does not change and 
neither a bound pole nor a virtual pole very close to the threshold 
appears.
This indicates that the structures in the spectra are essentially 
caused by the kinematical effects,
irrespective of the dynamical details.
The result also partly supports our assumption that the complex couplings
(fitting parameters)
well absorb 
the $Y_c\bar{D}^{(*)}$ coupled-channel effects 
that determine
the magnitude and phase of each one-loop amplitude.
Conversely, 
the $Y_c\bar{D}^{(*)}$ coupled-channel effects 
may be extracted from the data only if we understand the initial weak
vertices in advance.
Thus, to address the question at the beginning of
this paragraph, 
we need 
an experimental and/or lattice QCD determination of 
the initial weak vertices 
as well as 
higher statistics data for $B_s^0\to J/\psi p\bar{p}$.

The $P_c(4312)^+$ and $P_c(4337)^+$ peaks appear in 
$\Lambda_b^0\to J/\psi pK^-$ and $B_s^0\to J/\psi p \bar{p}$,
respectively, but not vice versa or both. 
Considering (anomalous) threshold cusps,
we can draw a consistent picture 
of this mutual exclusiveness
as follows.
As shown in Fig.~\ref{fig:comp-data}, the $P_c(4337)^+$ peak can be understood
as the $\Sigma_c\bar{D}$ threshold cusp. 
This may imply a comparable 
$\Lambda_c^+\bar{D}^{*0}$ threshold cusp, which is also supported
by the oscillating $\cos\theta_p$ distribution data in
Fig.~\ref{fig:comp-data-pp}(b).
In Refs.~\cite{DTS,DTS-pos},
the $P_c(4312)^+$ peak is explained by an interference between 
the $\Lambda_c^+\bar{D}^{*0}$ threshold cusp and
an anomalous $\Sigma_c\bar{D}$ threshold cusp
due to the double-triangle diagram. 
Thus, the $P_c(4312)^+$ and $P_c(4337)^+$ peak structures are consequences
of the different interference patterns of 
the (anomalous) $\Sigma_c\bar{D}$ and $\Lambda_c^+\bar{D}^{*0}$ threshold cusps.

In Ref.~\cite{bs0_beihang},
the authors discussed 
$P_c(4312)^+$ and $P_c(4337)^+$
as resonance or bound state poles from 
$\Lambda_c^+\bar{D}^{*0}-\Sigma_c^{(*)}\bar{D}$ 
coupled-channel dynamics. 
Within this picture, however, it is 
difficult to explain the above 
mutual exclusiveness of the $P_c^+$'s appearances.
Another possibility discussed in the paper
is that 
$P_c(4337)^+$ is a $\chi_{c0}\,p$ bound state (hadrocharmonium).
However, a strongly attractive charmonium-nucleon interaction 
does not seem likely since
lattice QCD calculations found rather weak
$J/\psi N$ and $\eta_c N$ interactions~\cite{jpsi-n-lqcd,sugiura}.
Also, this scenario implies that
the strength of a weak $B_s^0\to \chi_{c0}\,p\bar{p}$ decay vertex
would be significantly larger than that of 
$B_s^0\to J/\psi p\bar{p}$; this implication
does not seem likely, either. 
As mentioned earlier, 
the weak $B_s^0\to J/\psi p\bar{p}$ decay vertex would play a minor role
since the $p\bar{p}$ threshold enhancement is absent;
other $B_s^0\to {\rm charmonium} + p\bar{p}$ are also expected to be minor.
Thus, the scenarios discussed in Ref.~\cite{bs0_beihang}
do not seem as comparably convincing as the one proposed in this work.

\section{summary}

We developed a model to analyze the recent LHCb data on 
$B_s^0\to J/\psi p\bar{p}$.
The $\Sigma_c\bar{D}$ one-loop mechanism in the model causes 
a threshold cusp that fits well 
the $P_c(4337)^+$ peak in the $J/\psi p$ invariant mass distribution data.
We also showed that 
an oscillating behavior in 
the proton helicity angle distribution data 
can be understood as 
an interference between 
the $\Sigma_c\bar{D}$ 
and $\Lambda_c\bar{D}^{*}$ 
threshold cusps.
Combining these results with our earlier analysis on $\Lambda_b^0\to J/\psi p K^-$,
we conclude that 
$P_c(4312)^+$ in $\Lambda_b^0\to J/\psi p K^-$ and 
$P_c(4337)^+$ in $B_s^0\to J/\psi p\bar{p}$
are due to different interference patterns between 
the $\Sigma_c\bar{D}$ 
and $\Lambda_c\bar{D}^{*}$ 
(anomalous) threshold cusps.
In this way, we can consistently understand why 
the $P_c(4312)^+$ and $P_c(4337)^+$ peaks appear 
in $\Lambda_b^0\to J/\psi p K^-$ and $B_s^0\to J/\psi p\bar{p}$,
respectively, but not vice versa or both. 
The conclusion may also call for the reconsideration of the existing
data that have been interpreted as exotic hadron signals.

\begin{acknowledgments}
We thank S. Sakai for useful discussions.
We also acknowledge J. Haidenbauer for useful discussions on $p\bar{p}$
 threshold enhancements.
This work is in part supported by 
National Natural Science Foundation of China (NSFC) under contracts 
U2032103 and 11625523, 
and also by
National Key Research and Development Program of China under Contracts
 2020YFA0406400 (S.X.N.).
The support is also from 
JSPS KAKENHI under Grant No. JP20K14478 (Y.Y.)
and from
Grants-in Aid for Scientific Research on
Innovative Areas under Grant No. 18H05407 (A.H.).
\end{acknowledgments}




\end{document}